\documentclass[aip,jcp,reprint,noshowkeys]{revtex4-1}
\usepackage{graphicx,float,dcolumn,bm,xcolor,microtype,multirow,amscd,amsmath,amssymb,amsfonts,physics,upgreek,pifont,wrapfig,enumitem,siunitx}
\usepackage[version=4]{mhchem}
\usepackage[utf8]{inputenc}
\usepackage[T1]{fontenc}

\AtBeginDocument{\nocite{achemso-control}}

\usepackage{hyperref}
\hypersetup{
    colorlinks=true,
    linkcolor=blue,
    filecolor=blue,      
    urlcolor=blue,	
    citecolor=blue
}

\usepackage{caption} 
\usepackage{subcaption}
\usepackage{tabularx}
\usepackage{cleveref}

\usepackage{float} 
\usepackage{mathtools}

\newcommand{\bluesuperscript}[1]{\textcolor{blue}{\textsuperscript{#1}}}

\newcommand{\UCAM}{Yusuf Hamied Department of Chemistry, University of Cambridge, Lensfield Road, Cambridge, CB2 1EW, U.K.}
\newcommand{\UOX}{Department of Chemistry, Physical and Theoretical Chemical Laboratory, University of Oxford, South Parks Road, Oxford, OX1 3QZ, U.K.}
\newcommand{\EPFL}{Laboratory of Computational Science and Modeling, Institut des Mat\'eriaux, \'Ecole Polytechnique F\'ed\'erale de Lausanne, 1015 Lausanne, Switzerland}

\begin{document}

\title{Excited state-specific CASSCF theory for the torsion of ethylene} 
\author{Sandra~Saade}
\altaffiliation[Current Address: ]{\EPFL}
\affiliation{\UOX}
\affiliation{\UCAM}

\author{Hugh~G.~A.~Burton}
\email{hgaburton@gmail.com}
\affiliation{\UOX}
\affiliation{\UCAM}

\date{\today}

\begin{abstract}
\begin{wrapfigure}[11]{r}{0.40\textwidth}
    \flushleft
    \vspace{-0.4cm}
    \hspace{-1.7cm}
    \fbox{\includegraphics[width=0.40\textwidth]{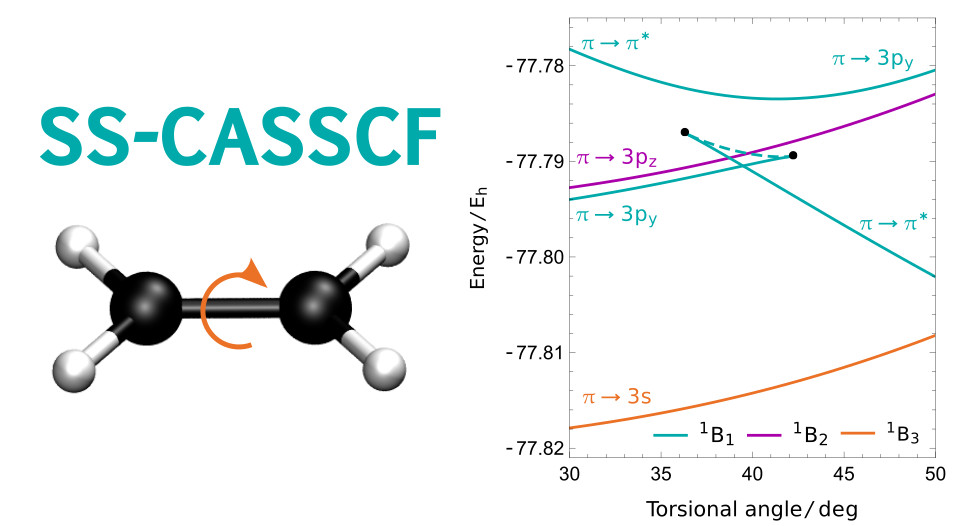}}
\end{wrapfigure}
State-specific complete active space self-consistent field (SS-CASSCF) theory has emerged as a promising route to accurately predict electronically excited energy surfaces  away from molecular equilibria. 
However, its accuracy and practicality for chemical systems of photochemical interest has yet to be fully determined.
We investigate the performance of SS-CASSCF theory for the low-lying ground and excited states in the double bond rotation of ethylene.
We show that state-specific approximations with a minimal (2e, 2o) active space  provide comparable accuracy to state-averaged calculations with much larger active spaces, while optimising the orbitals for each excited state significantly improves the spatial diffusivity of the wave function.
However, the unbalanced post-CASSCF dynamic correlation in valence and Rydberg excitations, or the use of a non-diffuse basis set, causes excited state solutions to coalesce and disappear, creating unphysical discontinuities in the potential energy surface. 
Our findings highlight the theoretical challenges that must be overcome to realise practical applications of state-specific electronic structure theory for computational photochemistry.
\end{abstract}

\maketitle

\section{Introduction}
Simulations of dynamic photochemical processes rely on faithful descriptions of ground- and excited-state
energy surfaces away from molecular equilibria, but obtaining accurate and efficient predictions of electronic excitations 
remains a major challenge.\cite{Gonzalez2011}
The prevalence of open-shell ground and excited states in photochemistry means that 
single-reference methods, such as equation-of-motion coupled cluster (EOM-CC)\cite{Bartlett2012}  and time-dependent 
density functional theory (TD-DFT),\cite{Dreuw2005} are generally restricted to  molecular structures around the equilibrium geometry.
Therefore, computational studies  rely on multi-configurational methods, 
usually in the form of state-averaged (SA) complete active space self-consistent-field (CASSCF) theory.\cite{Roos1980,RoosBook,Werner1981}
However, state-averaging can give discontinuous energy surfaces due to  ``root-flipping'' when electronic states cross,\cite{Zaitsevskii1994}
large active spaces are required to capture all the relevant states, 
and using a common set of orbitals does not account for bespoke orbital relaxation in charge transfer and Rydberg excitations.

Alternatively, recent research has explored the ``state-specific'' philosophy, where higher-energy electronic solutions  are used to 
approximate individual excited states, which formally exist as saddle points on the exact electronic energy landscape.\cite{Burton2021e}
The simplest approximation is self-consistent field (SCF) theory, where each excited state is represented by a single 
Slater determinant and the optimal orbitals are computed with either 
Hartree--Fock (HF) theory or Kohn--Sham density functional theory (KS-DFT).%
\cite{Gilbert2008,Besley2009,Barca2014,Barca2018,Hait2020,Hait2021,CarterFenk2020,Thom2008,Jensen2018,Burton2021a,Levi2020,Levi2020a}
This approach has proved to be successful for predicting double excitations, charge transfer
states\cite{Gilbert2008,Barca2018} and core excitations.\cite{Hait2020a}
However, for open-shell states away from the ground state equilibrium geometry, one must resort to symmetry-broken SCF
approximations that introduce spin or spatial symmetry contamination.\cite{Schmerwitz2022,Burton2018}
Furthermore, state-specific SCF solutions often disappear along a potential energy surface,\cite{Jensen2018,Burton2018,Vaucher2017,Burton2021e,Obermeyer2021} creating discontinuities that 
prevent dynamic simulations.

A more suitable approach for open-shell ground and excited states is 
 state-specific (SS) multi-configurational approximations, such as excited-state mean-field theory\cite{Shea2018,Hardikar2020} 
or CASSCF theory.\cite{Tran2019,Tran2020,Hanscam2022,Tran2023,Marie2023}
Compared to state-averaging, these approaches provide bespoke orbitals 
for each excitation, meaning that smaller active spaces can be used.\cite{Marie2023}
Using a minimal multi-configurational expansion to capture the key open-shell configurations is expected to alleviate the issues of disappearing SCF solutions.
Previous work has shown that unphysical solutions can still arise if the wrong active space is chosen,
and solutions can undergo symmetry breaking or disappear as the molecular structure changes.\cite{Marie2023}
However, the prevalence and significance of these irregularities for excited energy surfaces in larger molecules and basis sets
of photochemical interest remain to be determined,  preventing a firm evaluation of the long-term viability of SS-CASSCF theory.

In this contribution, we assess the performance of SS-CASSCF theory for the 
excited states in the double bond torsion of ethylene, which have been the subject of numerous  theoretical and experimental
studies over the past 50 years (see Ref.~\onlinecite{Feller2014} for an excellent overview).
The low-lying states of interest include the singlet and triplet
$\uppi\rightarrow 3\mathrm{s}$ and $\uppi\rightarrow 3\mathrm{p}$ Rydberg excitations, the $\pi \rightarrow \pi^*$ 
single excitation (V), and the $(\pi)^2 \rightarrow (\pi^*)^2$ double excitation (Z).
In particular, there has been significant debate about whether the $\pi \rightarrow \pi^*$ state has valence or Rydberg
character,\cite{Ryu1995,Huzinaga1962,Buenker1971,Dunning1969,Serrano1993,Sunil1984,McMurchie1977,Brooks1978,Angeli2009,Muller1999,Lindh1989,Feller2014}
which is compounded by the near degeneracy of the Rydberg and V single excitations and the
non-vertical nature of the experimental excitation.\cite{Johnson1979,Basch1970,Peyerimhoff1972,Petrongolo1983,Daday2012}
SA-CASSCF theory predicts the V state to be too diffuse in character,\cite{Serrano1993,Lindh1989,Muller1999}
 as measured by the spatial second-order moment $\ev*{x^2}$, which
is commonly attributed to the lack of dynamic correlation.\cite{McMurchie1977,Brooks1978,Sunil1984}
Furthermore, Angeli has highlighted the importance of dynamic $\upsigma$-polarisation and subsequent orbital contraction in the V state.\cite{Angeli2009}

At the planar $\mathrm{D_{2h}}$ structure, the bonding $\uppi$ and antibonding $\uppi^*$ orbitals transform as 
$\mathrm{b_{3u}}$ and $\mathrm{b_{2g}}$, respectively, 
where the \ce{C-C} bond coincides with the $z$-axis and the molecule lies in the $yz$-plane. The ground state and $\uppi \rightarrow \uppi^*$ open-shell singlet excitation correspond to the $1\,\mathrm{^1A_g}$ and 
$1\,\mathrm{^1B_{1u}}$ states.
Following a photoexcitation to the $1\,\mathrm{^1B_{1u}}$ state, the molecule is believed to rotate around the 
\ce{C-C} bond towards the twisted $\mathrm{D_{2d}}$ structure, before a further pyramidalization of a  $\ce{-CH2}$ group
leads to a conical intersection with the ground state.\cite{BenNun1998,BenNun2000,Barbatti2004}
Accurate excited-state energies along this torsional mode are therefore essential, but SA-CASSCF is susceptible to root-flipping.\cite{BenNun2000}
Since each  state is dominated by at most two determinants, we expect a state-specific
(2e, 2o) active space to give a qualitatively correct description.

In this work, we investigate the applicability of the 
SS-CASSCF\,(2,2) approach for the ground and excited states in the torsion of ethylene.
We show that multiple ground state solutions can occur, and we identify suitable stationary points for the 
low-lying Rydberg excitations, and the V and Z excited states.
We find that SS-CASSCF\,(2,2) can provide similar accuracy to SA-CASSCF calculations with much larger active spaces and
can recover the correct diffusivity of the valence V state.
On the other hand, we show that the incorrect ordering of excitations due to missing dynamic correlation or non-diffuse basis functions
can cause  solutions to disappear, giving unphysical energy surfaces.
Our findings highlight the promises and pitfalls of SS-CASSCF for practical excited-state applications.

\section{Theory}
\subsection{\label{methods}State-specific CASSCF theory}
Electronic states with unpaired electrons are inherently multi-configurational and must be modelled as a superposition of multiple
Slater determinants using configuration interaction (CI).
The complete active space (CAS) approach is the most common way to choose the subset of dominant configurations 
required to capture this ``static'' electron correlation.
In CASCI, a subset of relevant active orbitals are chosen and
a CI expansion is built using every possible way of arranging the active electrons in these partially occupied orbitals.
The remaining inactive and virtual orbitals are fully occupied, and empty, respectively, in each configuration.\cite{Seigbahn1981,Roos1980}
As a truncated CI expansion, the CASCI wave function depends strongly on the choice of orbitals in the inactive, active, and virtual spaces.
Therefore, the optimal wave function is usually identified by optimising the orbital and CI coefficients self-consistently with
the CASSCF approach.\cite{Roos1980}

On each optimisation step, the CASCI wave function is defined as
\begin{equation}
\ket{\Psi_J} = \sum_{I} \ket{\Phi_I} C_{IJ} ,
\end{equation}
where $C_{IJ}$ are the CI expansion coefficients for state $J$ in terms of the active Slater determinants $\ket{\Phi_I}$.
Variations in the CI and orbital coefficients can be represented using an exponential parametrisation as
\begin{equation}
	\ket*{\tilde{\Psi}_J} = \mathrm{e}^{\hat{R}}  \mathrm{e}^{\hat{S}} \ket {\Psi_{J}}.
\end{equation}
The anti-Hermitian operator $\hat{R}$ performs orbital rotations and is expressed in terms of the current orbitals\cite{Dalgaard1979,Dalgaard1978,Yeager1979} 
\begin{equation}
	\hat{R} = \sum_{p>q} R_{pq} {\hat{E}^{-}}_{pq} 
\end{equation}
where ${\hat{E}^{-}}_{pq} = \sum_{\sigma \epsilon {\uparrow, \downarrow}} {\hat{a}_{q\sigma}}^{\dagger} \hat{a}_{p\sigma} - {\hat{a}_{p\sigma}}^{\dagger} \hat{a}_{q\sigma}$ is the anti-Hermitian singlet excitation operator.\cite{HelgakerBook}
Similarly, the $\hat{S}$ operator transforms the CI expansion by considering the transfer operators between the 
target state $\ket {\Psi_{J}}$ and the orthogonal states $\ket {\Psi_{K}}$ in the CASCI space\cite{Yeager1979},
\begin{equation}
	\hat{S} = \sum_{K\neq J} S_{K}\, \Big(\dyad{\Psi_K}{\Psi_J}  - \dyad{\Psi_J}{\Psi_K} \Big).
\end{equation}
The  energy $E_J(\boldsymbol{R, S}) = \bra {\Psi_J} \mathrm{e}^{-\hat{S}} \mathrm{e}^{-\hat{R}} \hat{H} \mathrm{e}^{\hat{R}} \mathrm{e}^{\hat{S}} \ket{\Psi_J}$ is then a function of the variables $S_K$ and $R_{pq}$ and the optimal CASSCF solutions
are stationary points on the corresponding electronic energy landscape. 

\subsection{Computational details}
Since exact excited states are higher-index 
saddle points of the electronic energy landscape,\cite{Burton2021e} we expect
SS-CASSCF excited states to also be saddle points of the energy.
These can be  identified using second-order
optimisation schemes, which also accelerate convergence if there
is strong coupling between the orbital and CI degrees of
freedom.\cite{Yeager1979,Das1973,Yeager1980a,Yeager1980b}
We employ the eigenvector-following technique\cite{WalesBook} to target stationary points 
with a particular Hessian index, as described in Ref.~\onlinecite{Marie2023}.
For open-shell single excitations, an initial guess can be prepared by first optimising 
the orbitals for a suitable configuration state function (CSF) following the framework outlined in Ref.~\onlinecite{HelgakerBook}.
Once an optimal SS-CASSCF solution has been found, it can be used as an initial guess for the 
next molecular geometry, allowing it to be tracked across the full potential energy surface.
Since the Hessian index may change along a binding curve, the mode-controlled 
Newton--Raphson optimiser described in Ref.~\onlinecite{Marie2023} 
is used to reconverge solutions at each geometry without prior knowledge of the Hessian index.

All calculations are performed using an in-house computational package developed in our group, which 
forms an extension to \textsc{PySCF}.\cite{PySCFb}
We consider the aug-cc-pVDZ basis set,\cite{Dunning1989,Kendall1992} which includes support for diffuse Rydberg states, and the smaller 6-31G basis set.\cite{Ditchfield1971} 
The convergence threshold is set to a root-mean-square gradient value of $10^{-7}\,\mathrm{E_h}$.
Figures are plotted using Mathematica 12.0\cite{Mathematica} and orbitals are visualised using the VMD software.\cite{vmd}

\section{Results and Discussion}

\subsection{Summary of SS-CASSCF\,(2,2) solutions}

\begin{figure}[htb]
\includegraphics[width=\linewidth]{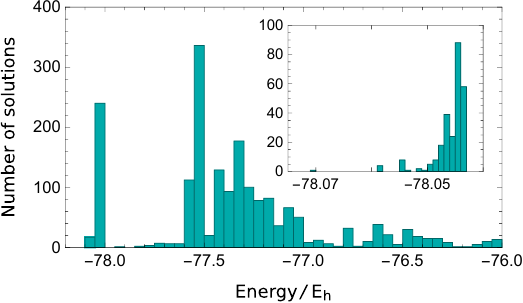}
\caption{Number of SS-CASSCF\,(2,2) solutions identified at the $\mathrm{D_{2h}}$ geometry (aug-cc-pVDZ)  using random starting guesses. 
\textit{Inset:} The number of stationary points associated with the closed-shell ground state.}
\label{fig:histo}
\end{figure}

Using the aug-cc-pVDZ basis set, we first characterised the SS-CASSCF\,(2,2) solution space
by starting from random MO and CI coefficients.
We considered the planar $\mathrm{D_{2h}}$ geometry used in Ref.~\onlinecite{Loos2018b}, which is 
provided in the \textcolor{blue}{Supporting Information}.
Low-energy solutions were targetted by searching for stationary points 
with Hessian indices between 0--10 using  eigenvector-following.
Up to 1000 random starting points were tested for each Hessian index.
An extremely large number of low-energy solutions were identified, as illustrated in Fig.~\ref{fig:histo},
making a complete characterisation of the solution space impossible.

\begin{figure*}[htb!]
\includegraphics[width=0.8\linewidth]{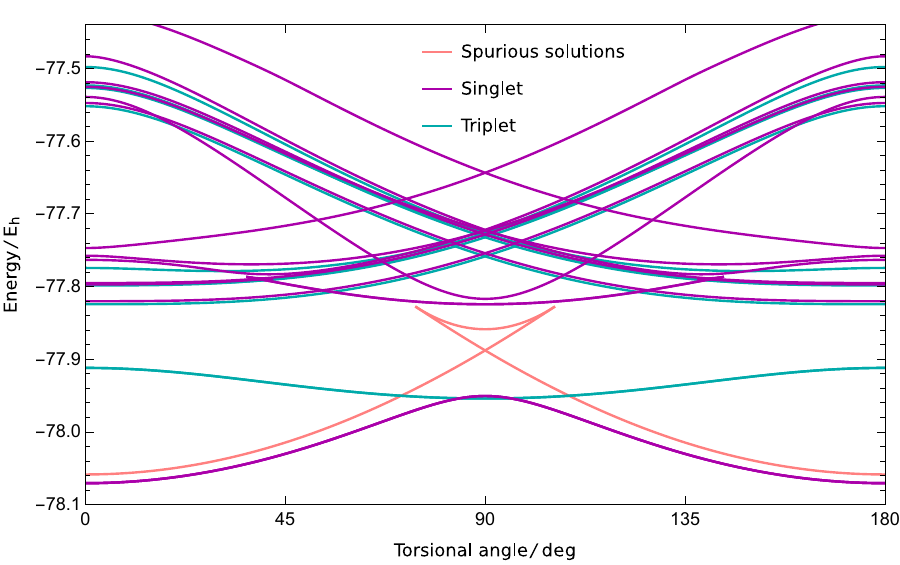}
\caption{Summary of the physically meaningful singlet and triplet SS-CASSCF\,(2,2) solutions in ethylene (aug-cc-pVDZ),
as well as the spurious local minima and index-1 saddle point (see Fig.~\ref{fig:new_fig2}\textcolor{blue}{C}).}
\label{fig:new_fig1}
\end{figure*}

Instead, we focussed our attention on the solutions corresponding to local minima,  the low-energy singlet and triplet single excitations, 
and the Z double excitation.
Starting from a pre-optimised open-shell CSF allowed 
suitable stationary points to be found for the $(\pi \rightarrow \mathrm{3s})$,  $(\pi \rightarrow \mathrm{3p})$,  
$(\pi \rightarrow \pi^*)$ excitations, among others.
The $(\pi)^2 \rightarrow (\pi^*)^2$ double excitation 
was identified by starting at the corresponding non-aufbau Slater determinant.
Tracing the relevant solutions across the double bond rotation resulted in the ground- and excited-state
energy surfaces shown in Fig.~\ref{fig:new_fig1}.

Some solutions disappear along the torsional rotation.
This disappearance can only occur if two stationary points coalesce  on the 
CASSCF energy landscape at a pair annihilation point,\cite{Burton2018,Marie2023}
which mathematically corresponds to a fold catastrophe.\cite{GilmoreBook}
This coalescence  is associated with the onset of a zero  eigenvalue in the Hessian matrix of second derivatives with respect to the wave function parameters, and similar phenomena occur for multiple Hartree--Fock solutions.\cite{Burton2018,Marie2023,Burton2021a,Huynh2019}
The other solution involved in the pair annihilation can be identified using a line search in the
direction of the eigenvector corresponding to the zero Hessian 
eigenvalue, as detailed in Appendix~\ref{sec:linesearch}.

In the following Sections, we characterise the local minima (Section~\ref{sec:minima}) and
the valence and Rydberg excitations (Section~\ref{sec:valryd}). Finally, we highlight how the SS-CASSCF solutions change if we use a smaller basis set that cannot describe Rydberg states (Section~\ref{sec:basis}).

\subsection{Multiple local minima}
\label{sec:minima}
\begin{figure*}[htb]
	\includegraphics[width=0.8\linewidth]{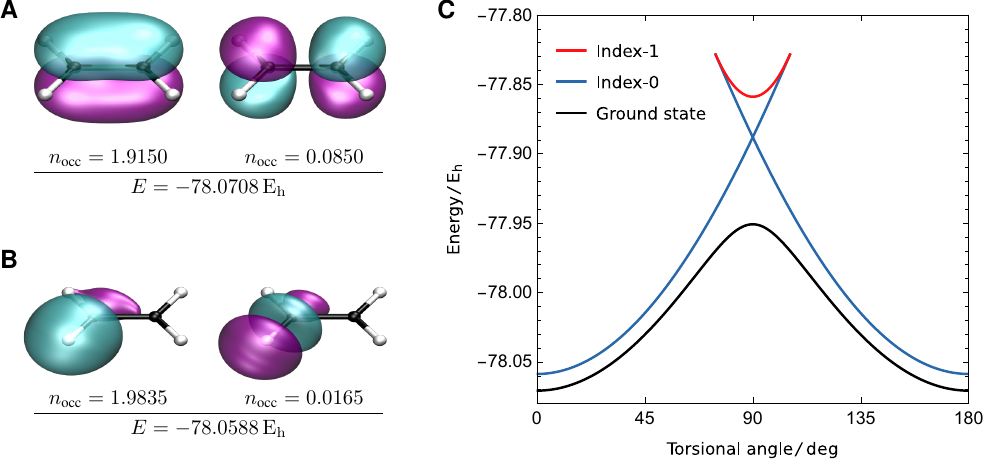}
	\caption{Comparison of the natural active orbitals for the SS-CASSCF\,(2, 2) minima at the planar geometry (aug-cc-pVDZ).
The global minimum (\textbf{\textsf{A}}) gives a smooth torsional barrier,
while the local minima (\textbf{\textsf{B}}) give a cusp 
at $\SI{90}{\deg}$ and disappear in a pair annihilation point at $\SI{106}{\deg}$.}
	\label{fig:new_fig2}
\end{figure*}

Although there is only one minimum on the exact energy landscape,\cite{Burton2021e} the 
SS-CASSCF\,(2,2) approximation yields five minima at the planar structure, 
corresponding to a unique global minimum and a four-fold degenerate set of local minima.
The partially occupied natural orbitals for these solutions reveal that the global minimum corresponds to the 
expected $\{\uppi, \uppi^* \}$ active orbitals with occupations of 1.9150 and 0.0850, 
respectively (Fig.~\ref{fig:new_fig2}\textcolor{blue}{A}).
In contrast, the active orbitals for the local minima break the spatial symmetry 
and correspond to the quasi-localised \ce{C-H} $\upsigma$ and $\upsigma^*$ orbitals, with the four-fold degeneracy 
arising from the four \ce{C-H} bonds (Fig.~\ref{fig:new_fig2}\textcolor{blue}{B}).
Since the true ground state is dominated by one closed-shell configuration, both active spaces include one orbital 
that is almost doubly occupied and one that is almost unoccupied.
The active orbital with $n_\text{occ} \approx 2$ can be swapped for a doubly occupied inactive orbital 
without significantly changing the energy, leading to multiple representations of the ground state,  as described in Ref.~\onlinecite{Marie2023}.
Therefore, in the absence of strong static correlation at the $\mathrm{D_{2h}}$ geometry, the different minima attempt to capture dynamic correlation in either the 
\ce{C-H} $\upsigma$ or \ce{C-C} $\uppi$ bonds.

Although both sets of minima provide a reasonable approximation to the planar geometry,  
choosing the right active orbitals is essential for computing physically meaningful energy surfaces.\cite{Marie2023}
The global minimum can be followed across the full torsion to give a smooth rotational barrier 
(Fig.~\ref{fig:new_fig2}\textcolor{blue}{C}) because the $\{\uppi, \uppi^* \}$ active orbitals can correctly break 
the \ce{C-C} $\uppi$ bond.
In contrast, the energy of the \ce{C-H} $\{\upsigma, \upsigma^{*} \}$  local minimum does not reach a maximum at  $\SI{90}{\deg}$, 
and the solution eventually coalesces with an index-1 saddle point, both of which disappear in a pair annihilation point at $\SI{106}{\deg}$. 
The corresponding index-1 saddle point can be traced from $\SI{106}{\deg}$ back to $\SI{74}{\deg}$, where it coalesces with a 
symmetry-related \ce{C-H} $\{\upsigma, \upsigma^{*} \}$  local minimum that can be identified at the $\SI{180}{\deg}$ planar structure. 
This coalescence pattern between symmetry-related local minima and a connecting index-1 saddle point is a common feature 
of non-linear wave function approximations.\cite{Burton2018,Huynh2019,Marie2023} 
Its presence for the local SS-CASSCF\,(2,2) minima in ethylene  re-emphasises the importance
of selecting meaningful active spaces that can faithfully capture the static correlation across a particular chemical reaction coordinate.

\subsection{Valence and Rydberg excitations}
\label{sec:valryd}
The low-lying singly excited states in ethylene correspond to excitations from the $\uppi$ orbital 
to a $3\mathrm{s}$ or $3\mathrm{p}$ Rydberg orbital, and the valence $\uppi\rightarrow \uppi^*$ excitation.
A SS-CASSCF\,(2,2) solution for each of the corresponding singlet and triplet excitations 
can be identified at the planar geometry.
The orbital assignment and excitation energies are tabulated in Table~\ref{tab:sing_excitation}, 
alongside literature benchmark values.
Remarkably, the SS-CASSCF\,(2,2) excitation energies correspond closely to SA-CASSCF results
computed with a much larger (2,11) active space.\cite{Serrano1993} 
This result highlights that the state-specific approach can provide accurate energies with
significantly smaller active spaces by considering only the active orbitals that are directly involved in the excitation.

\begin{table}[htb]
\caption{Vertical excitation energies ($\mathrm{eV}$) computed with SS- and  SA-CASSCF are compared against theoretical best estimates (TBE).}
\label{tab:sing_excitation}
\begin{ruledtabular}
\begin{tabular}{llcccc}
State & & SS-(2,2)\bluesuperscript{a} & SA-(2,11)\bluesuperscript{b} & TBE\bluesuperscript{c}
\\
\hline
$\mathrm{1\,^1A_g}$  &                                             & $\hphantom{-}0.00$ & $\hphantom{-}0.00$  & $0.00$ \\
$\mathrm{1\,^1B_{1u}}$ &  $\mathrm{\uppi\rightarrow\uppi^*}$       & $\hphantom{-}8.36$ & $\hphantom{-}8.20$  &  $8.00$ \\
$\mathrm{1\,^1B_{3u}}$ &  $\mathrm{\uppi\rightarrow3s}$            & $\hphantom{-}6.81$ & $\hphantom{-}6.82$  &  $7.45$ \\
$\mathrm{1\,^1B_{1g}}$ &  $\mathrm{\uppi\rightarrow3p_y}$          & $\hphantom{-}7.44$ & $\hphantom{-}7.43$  & $8.06$ \\
$\mathrm{1\,^1B_{2g}}$ &  $\mathrm{\uppi\rightarrow3p_z}$          & $\hphantom{-}7.49$ & $\hphantom{-}7.51$  & $8.11$ \\
\hline
$\mathrm{1\,^3B_{1u}}$ &  $\mathrm{\uppi\rightarrow\uppi^*}$       & $\hphantom{-}4.32$ & $\hphantom{-}4.65$   &   $4.55$ \\
$\mathrm{1\,^3B_{3u}}$ &  $\mathrm{\uppi\rightarrow3s}$            & $\hphantom{-}6.70$ & $\hphantom{-}6.74$   &    $7.29$ \\
$\mathrm{1\,^3B_{1g}}$ &  $\mathrm{\uppi\rightarrow3p_y}$          & $\hphantom{-}7.40$ & $\hphantom{-}7.41$   & $8.02$ \\
$\mathrm{1\,^3B_{2g}}$ &  $\mathrm{\uppi\rightarrow3p_z}$          & $\hphantom{-}7.43$ & $\hphantom{-}7.57$   & $8.04$ \\ \hline
\multirow{2}{*}{MUE} & Rydberg & $-0.62$            & $-0.60$                 & --
\\                   & Valence & $\hphantom{-}0.06$ & $\hphantom{-}0.15$      & --
\\ \hline
\multirow{2}{*}{MAE} & Rydberg & $\hphantom{-}0.62$ & $\hphantom{-}0.60$ &  --
\\                   & Valence & $\hphantom{-}0.29$ & $\hphantom{-}0.15$ &  --

\end{tabular}
\end{ruledtabular}
\footnotesize{(a) This work,
(b) Ref.~\onlinecite{Serrano1993},
(c) Ref.~\onlinecite{Feller2014}.}
\end{table}

Compared to the theoretical best estimates (TBE) from Ref.~\onlinecite{Feller2014}, the SS-CASSCF\,(2,2) Rydberg excitation energies 
are consistently underestimated by around $\SI{0.6}{\eV}$, as shown by the mean unsigned error (MUE) 
in Table~\ref{tab:sing_excitation}.
Since the SS-CASSCF approximation predominantly captures static electron correlation, 
this consistent shift suggests that there is an imbalance between the dynamic correlation 
in the ground and Rydberg states, supporting the findings of Ref.~\onlinecite{Serrano1993}.
In particular, the spatially compact nature of the ground state leads to regions of higher electron density and thus greater dynamic correlation 
than the more diffuse Rydberg states. 
Therefore, SS-CASSCF\,(2,2) underestimates the ground-state  energy and, by extension, the Rydberg  excitation energies.
The second-order moment $\ev*{x^2}$ for the Rydberg states have an error around 
$2.5\,\mathrm{a_0^2}$, as shown in Table~\ref{tab:singles}, indicating that the state-specific wave functions are at least qualitatively accurate.

\begin{figure}[htb]
\includegraphics[width=\linewidth]{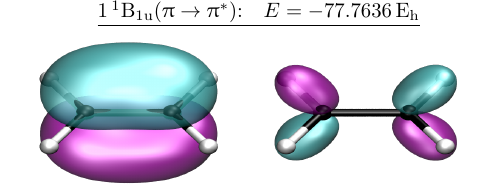}
\caption{HOMO and LUMO orbitals for the $1\,\mathrm{^1B_{1u}}\ (\uppi\rightarrow\uppi^*)$ excitation in planar ethylene. 
The $\uppi^*$ orbital is significantly contracted compared to Fig.~\ref{fig:new_fig2}\textcolor{blue}{A}.} 
\label{fig:new_fig3}
\end{figure}

Whether the $1\,\mathrm{^1B_{1u}}$ $\uppi \rightarrow \uppi^*$ state has predominant valence or Rydberg character
has long been disputed due to the challenge of  
reproducing the experimental band absorption maximum 
at $\SI{7.6}{\eV}$.
Recent studies have confirmed that nonadiabatic effects\cite{Peyerimhoff1972,Buenker1971,Petrongolo1983}
shift this experimental value from the vertical 
excitation energy that is closer to $\SI{8.0}{\eV}$.\cite{Feller2014,Loos2018b,Daday2012}
The degree of Rydberg character can be measured through the $\ev*{x^2}$ value, which 
can vary significantly for a small change in energy,\cite{Brooks1978}
while dynamic correlation and $\upsigma$-polarisation are expected to cause the
excited state $\pi^*$ orbital to contract.\cite{Angeli2009}
This spatial contraction is not seen in state-averaged CASSCF,\cite{Angeli2009,Muller1999,Lindh1989} 
as shown by the large $\ev{x^2}$ value for SA-CASSCF\,(2,11)\cite{Serrano1993} in Table~\ref{tab:singles}.
In contrast, the SS-CASSCF\,(2,2) excited state clearly yields a more contracted $\pi^*$ orbital (Fig.~\ref{fig:new_fig3})
than the ground state solution, giving a $\ev{x^2}$ value that closely matches the TBE.

The improvement in the spatial diffusivity of the SS-CASSCF wave function is also reflected in the oscillator 
strength for the $\uppi \rightarrow \uppi^*$ excitation.
Since the state-specific  ground and excited states are represented with different sets of orbitals, we use the 
extended nonorthogonal Wick's theorem\cite{Burton2021c,Burton2022} implemented in the
 \textsc{LibGNME} software package\cite{BurtonGNME} to evaluate the transition dipole moment.
Compared to the $0.17\,\mathrm{a.u.}$ error for the SA-CASSCF\,(2,11) result from Ref.~\onlinecite{Serrano1993}, the 
SS-CASSCF\,(2,2) approach  predicts the oscillator strength with a deviation of $0.035\,\mathrm{a.u.}$ from the TBE (Table~\ref{tab:singles}).
This improvement suggests that the state-specific approach reduces the contamination from 
nearby Rydberg states, which have a weaker oscillator strength than the valence excitation.

\begin{table*}[htb]
\caption{Comparison of the vertical excitation energy $\Delta E$ ($\mathrm{eV}$), oscillator strength $f$ ($\mathrm{a.u.}$)
and second-order moment $\ev{x^2}$ ($\mathrm{a_0^2}$) for SS-CASSCF\,(2,2) and SA-CASSCF\,(2,11) at the planar $\mathrm{D_{2h}}$
geometry (aug-cc-pVDZ).
The state-specific approach significantly improves the $\ev*{x^2}$  value for the valence $\uppi \rightarrow \uppi^*$
state compared to SA-CASSCF. }
\label{tab:singles}
\begin{ruledtabular}
\begin{tabular}{llccccccccc}
& & \multicolumn{3}{c}{SS-CASSCF\,(2,2)\bluesuperscript{a}} & \multicolumn{3}{c}{SA-CASSCF\,(2,11)\bluesuperscript{b}}  & \multicolumn{3}{c}{TBE\bluesuperscript{c}} 
\\ \cline{3-5} \cline{6-8} \cline{9-11}
\multicolumn{2}{l}{State} & $\Delta E$ &$f$  & $\ev*{x^2}$ & $\Delta E$ &$f$  & $\ev*{x^2}$ & $\Delta E$ &$f$  & $\ev*{x^2}$
\\
\hline
$\mathrm{1\,^1A_g}$  &                                       & $0.00$ &         & $11.68$  & $0.00$ &         & $11.8$ & $0.00$ &         & $11.78$  \\
$\mathrm{1\,^1B_{1u}}$ &  $\mathrm{\uppi\rightarrow\uppi^*}$ & $8.36$ & $0.298$ & $17.89$  & $8.20$ & $0.16\hphantom{0}$  & $44.1$ & $8.00$ & $0.333$ & $17\pm1$ \\
$\mathrm{1\,^1B_{3u}}$ &  $\mathrm{\uppi\rightarrow3s}$      & $6.81$ & $0.066$ & $21.55$  & $6.82$ & $0.067$ & $24.3$ & $7.45$ & $0.069$ & $23.96$  \\
$\mathrm{1\,^1B_{1g}}$ &  $\mathrm{\uppi\rightarrow3p_y}$    & $7.44$ &         & $17.89$  & $7.43$ &         & $17.2$ & $8.06$ &         & $20.38$  \\
$\mathrm{1\,^1B_{2g}}$ &  $\mathrm{\uppi\rightarrow3p_z}$    & $7.49$ &         & $18.84$  & $7.51$ &         & $18.0$ & $8.11$ &         & $21.53$  \\
\hline
$\mathrm{1\,^3B_{1u}}$ &  $\mathrm{\uppi\rightarrow\uppi^*}$ & $4.32$ &         & $11.74$  & $4.65$ &         & $11.9$ & $4.55$ &         & $11.69$  \\
$\mathrm{1\,^3B_{3u}}$ &  $\mathrm{\uppi\rightarrow3s}$      & $6.70$ &         & $21.40$  & $6.74$ &         & $23.8$ & $7.29$ &         & $23.45$  \\
$\mathrm{1\,^3B_{1g}}$ &  $\mathrm{\uppi\rightarrow3p_y}$    & $7.40$ &         & $17.68$  & $7.41$ &         & $17.0$ & $8.02$ &         & $19.66$  \\
$\mathrm{1\,^3B_{2g}}$ &  $\mathrm{\uppi\rightarrow3p_z}$    & $7.43$ &         & $18.48$  & $7.47$ &         & $17.7$ & $8.04$ &         & $20.35$  
\end{tabular}
\end{ruledtabular}
\footnotesize{(a) This work,
(b) Ref.~\onlinecite{Serrano1993},
(c) Ref.~\onlinecite{Feller2014}.}
\end{table*}

\begin{figure}[b]
\includegraphics[width=\linewidth]{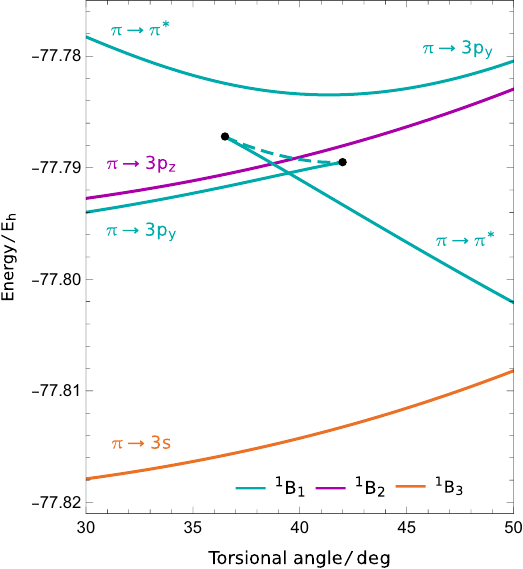}
\caption{SS-CASSCF\,(2,2) predicts the wrong ordering for the $\pi \rightarrow \pi^*$ and  $\mathrm{\uppi\rightarrow3p_y}$ states
at the planar geometry, leading to an avoided crossing along the torsional rotation. The lower energy solution 
disappears at a pair annihilation point ($\SI{42}{\deg}$) and a new discontinuous SS-CASSCF\,(2,2) solution emerges
($\SI{37}{\deg}$), which represents the $\pi \rightarrow \pi^*$ state at larger torsional angles. 
Rydberg states with different symmetries are unaffected.}
\label{fig:new_fig4}
\end{figure}

Compared to the $\SI{-0.6}{\eV}$ underestimate for  the Rydberg excitation energies, the SS-CASSCF\,(2,2) approximation
overestimates the V excitation energy by $\SI{0.36}{\eV}$. 
This overestimate can be understood because the  $\pi \rightarrow \pi^*$ is dominated by zwitterionic resonance
structures with a larger dynamic correlation energy than the ground state, 
which is not captured by the CASSCF approximation.\cite{Sunil1984,McMurchie1977,Angeli2009}
Therefore, the SS-CASSCF\,(2,2) approximation erroneously predicts that the valence $1\,\mathrm{^1B_{1u}}$ state is higher in 
energy than the Rydberg $1\,\mathrm{^1B_{1g}}$ and  $1\,\mathrm{^1B_{2g}}$ states at the planar geometry.
This incorrect ordering has profound consequences on the corresponding excited-state energy surfaces along the torsional rotation.
As the molecule twists away from the planar geometry, the spatial point group changes from $\mathrm{D_{2h}}$ to $\mathrm{D_{2d}}$.
Under this descent in symmetry, the planar  $1\,\mathrm{^1B_{1u}}$ and $1\,\mathrm{^1B_{1g}}$ both transform as the same 
$\mathrm{^1B_1}$ irreducible representation, meaning that they can now couple through the Hamiltonian.
The $\mathrm{\uppi\rightarrow\uppi^*}$ and  $\mathrm{\uppi\rightarrow3p_y}$ excited states become lower and higher in energy, 
respectively, eventually leading to an unphysical avoided crossing (cyan in Fig.~\ref{fig:new_fig4}).

State-specific approximations are known to have unphysical solutions or coalescence points in the vicinity of 
avoided crossings.\cite{Burton2018,Marie2023,Jensen2018}
Here, we see that the higher energy solution, corresponding to the planar $\mathrm{\uppi\rightarrow\uppi^*}$ state,
continuously transforms into the $\mathrm{\uppi\rightarrow3p_y}$ state at the avoided crossing, as shown in Fig.~\ref{fig:new_fig4}.
In contrast, the lower energy solution continues to increase in energy until it eventually disappears in a pairwise coalescence point 
at $\SI{42}{\deg}$.
The other solution involved in the coalescence can be followed back to $\SI{37}{\deg}$, where it coalesces with a third solution
that corresponds to the  $\mathrm{\uppi\rightarrow\uppi^*}$ state after the avoided crossing.
Therefore, the lower $\mathrm{\uppi\rightarrow3p_y}$ solution is the only physically meaningful state that 
cannot be followed across the full torsional rotation, creating potential issues for the use of SS-CASSCF theory in \textit{ab initio} excited-state molecular dynamics.
We attempted to avoid this issue using a (2e, 3o) active space that contained both the $\uppi^*$ and $\mathrm{3p_y}$ with no success.
On the other hand, the state-specific philosophy successfully avoids the more widespread discontinuities that occur in 
state-averaged calculations, as seen in Fig.~6 of Ref.~\onlinecite{BenNun2000}.
 
Finally, we consider the double excitation $(\uppi)^2\rightarrow (\uppi^*)^2$, which cannot be captured by linear response formalisms
such as TD-DFT.
Starting from the non-aufbau Slater determinant at the planar geometry, 
the corresponding  SS-CASSCF\,(2,2) solution can be identified with an excitation energy  of $\Delta E = \SI{14.46}{\eV}$ and 
provides a continuous energy surface across the full torsional rotation (Fig.~\ref{fig:new_fig1}).
This state is less well covered in the literature, but benchmark values from the QUEST dataset\cite{Loos2019,Veril2021} and 
Ref.~\onlinecite{Barbatti2004} predict an excitation energy closer to $13$--$\SI{13.6}{\eV}$.
Therefore, the SS-CASSCF\,(2,2) overestimates the double excitation energy, which we 
believe is a result of  the unbalanced dynamic correlation 
between the ground and excited states, as already seen for the $\uppi \rightarrow \uppi^*$ excitation.

\subsection{Consequences of a non-diffuse  basis set}
\label{sec:basis}

\begin{figure*}[htb]
\includegraphics[width=\linewidth]{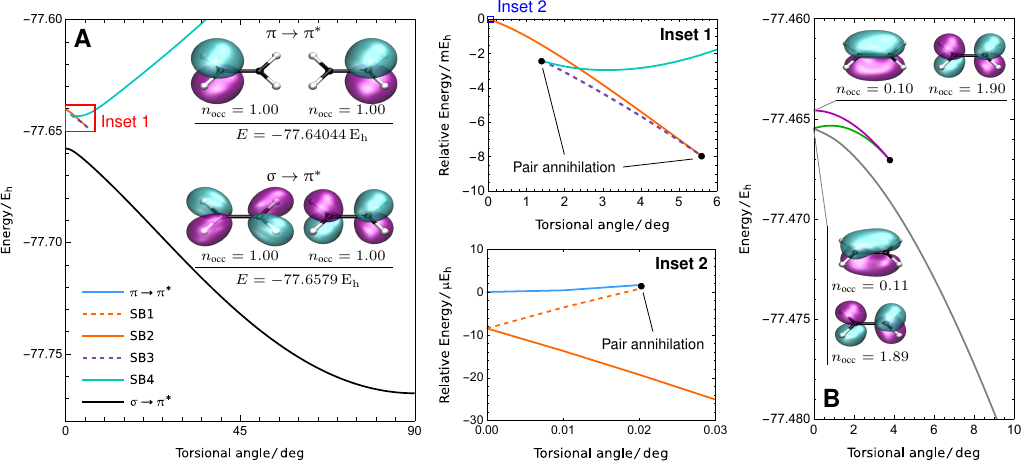}
\caption{SS-CASSCF\,(2,2) with the 6-31G basis set does not provide physically meaningful 
energy surfaces for the singly excited  $\uppi\rightarrow\uppi^*$  state or the doubly excited  $(\uppi)^2\rightarrow(\uppi^*)^2$ state.
(\textbf{\textsf{A}}) This approximation predicts the wrong ordering of the $\uppi\rightarrow\uppi^*$ 
and  $\upsigma\rightarrow\uppi^*$ states at the planar geometry, leading to a series of symmetry-broken (SB) solutions and an unphysical avoided crossing.
(\textbf{\textsf{B}}) The symmetry-pure solution (purple) corresponding to the $(\uppi)^2\rightarrow(\uppi^*)^2$ excitation
disappears at a torsional angle of $\SI{3.8}{\deg}$, giving an unphysical potential energy surface. }
\label{fig:basis}
\end{figure*}

The presence of low-energy Rydberg states means that diffuse basis functions are considered to be essential 
for accurately predicting the excited states in ethylene.\cite{Feller2014,Huzinaga1962,Dunning1969}
We also performed SS-CASSCF\,(2,2) calculations using the 6-31G basis set, highlighting how the lack of diffuse basis 
functions can fundamentally change the pattern of state-specific solutions in ethylene. 
While the ground state exhibited a global minimum and four-fold degenerate local minima that are directly analogous 
to the aug-cc-pVDZ basis, we were unable to find any physically meaningful approximations to the singly excited  $\uppi \rightarrow \uppi^*$ or the doubly excited $(\uppi)^2 \rightarrow (\uppi^*)^2$ energy surfaces.

To target the $\uppi \rightarrow \uppi^*$ excited state, we started the SS-CASSCF\,(2,2) optimisation from the output of  
a state-averaged CASSCF\,(2,2) calculation at the planar geometry.
The planar molecular structure was identified through a geometry optimisation using the B3LYP functional and is provided in the \textcolor{blue}{Supporting Information}. 
Starting from the state-averaged $\uppi \rightarrow \uppi^*$ initial guess gave a stationary point with symmetry-pure orbitals, with the natural orbitals corresponding to the localised zwitterionic configurations (Fig.~\ref{fig:basis}\textcolor{blue}{A}).
However, this solution only exists up to a torsional angle of $\SI{0.02}{\deg}$, where it disappears in a pair annihilation point 
(Fig.~\ref{fig:basis}: Inset 2). 
A complex pattern of coalescing solutions can be found that ultimately connects the $\uppi \rightarrow \uppi^*$ solution 
to another solution that emerges at $\SI{1.5}{\deg}$, which increases in energy for higher torsional angles (cyan in Fig.~\ref{fig:basis}\textcolor{blue}{A}).

Alternatively, searching for the  $\uppi \rightarrow \uppi^*$ state at $\SI{90}{\deg}$ yields a solution that 
exists all the way to $\SI{0}{\deg}$ (black in Fig.~\ref{fig:basis}). However, the corresponding natural orbitals at the planar geometry represent the $\mathrm{^1B_{1g}}$ $\upsigma \rightarrow \uppi^*$ 
excitation, which is known to be higher in energy than the $\uppi \rightarrow \uppi^*$ state.\cite{Feller2014}
Ultimately, the smaller 6-31G basis set results in the incorrect ordering of the  $\mathrm{^1B_{1g}}$  and  $\mathrm{^1B_{1u}}$ 
excited states because it cannot describe the diffuse character of the $\uppi \rightarrow \uppi^*$ state,
as indicated by the small $\ev{x^2}$ value of $12.24\,\mathrm{a_0^2}$. 
Like the interaction between the $\uppi \rightarrow \uppi^*$ excitation and the Rydberg states using the aug-cc-pVDZ basis, this
ordering problem creates an unphysical avoided crossing that causes SS-CASSCF\,(2,2) solutions to coalesce and disappear as the double bond rotates, leading to catastrophic potential energy surfaces.

Similarly, starting from the state-averaged states allows a symmetry-pure SS-CASSCF\,(2,2) solution  to be identified for the  $(\uppi)^2 \rightarrow (\uppi^*)^2$ double excitation (purple in Fig.~\ref{fig:basis}\textcolor{blue}{B}).
However, this solution also disappears as the molecule is twisted and cannot be traced beyond $\SI{3.8}{\deg}$, 
where it coalesces with another solution (green in Fig.~\ref{fig:basis}\textcolor{blue}{B}). 
This second state can be traced back to  the planar geometry, where it forms a pair of degenerate solutions with natural orbitals that break the spatial symmetry (the degeneracy is lifted for non-zero torsional angles). 
The other degenerate solution can be followed across the full torsional mode for angles between $0$ and $\SI{180}{\deg}$ (grey in Fig.~\ref{fig:basis}\textcolor{blue}{B}). 
However, as these degenerate solutions break the spatial symmetry and cross in energy at $\SI{0}{\deg}$, neither predicts a stationary point in the excited energy surface at the planar geometry.
Consequently, the SS-CASSCF\,(2,2) approximation is not able to provide any meaningful potential energy surface for the $(\uppi)^2 \rightarrow (\uppi^*)^2$ Z state of ethylene using the 6-31G basis, 
and it is vital that the basis set is sufficient for the excited states of interest.

\section{Concluding Remarks}

Excited state-specific approximations promise to overcome the challenges of state-averaged CASSCF theory for
predicting excited energy surfaces by facilitating calculations with smaller active spaces and 
avoiding root-flipping discontinuities.
In this work, we assessed the performance of the SS-CASSCF\,(2,2) approach for the valence and 
Rydberg excitations in the torsion of ethylene, using the aug-cc-pVDZ and 6-31G basis sets. 
While a large number of SS-CASSCF\,(2,2) solutions exist, we were able to target physically meaningful stationary points for the
low-lying excited states at the planar $\mathrm{D_{2h}}$ structure using the aug-cc-pVDZ basis set.
These solutions provided  excitation energies and properties that are comparable to much larger state-averaged approximations, highlighting that SS-CASSCF can be applied with only the active orbitals that are involved in each excitation.
Furthermore, most of the SS-CASSCF\,(2,2) solutions using aug-cc-pVDZ 
can be continuously followed across the torsional rotation, 
avoiding the root-flipping problems in SA-CASSCF
and the limitations of single-reference linear-response methods.

The imbalance between the missing dynamic correlation in Rydberg and valence excited states means that SS-CASSCF\,(2,2) theory fails to provide the correct state ordering  in planar ethylene.
This incorrect ordering of the $\mathrm{\uppi\rightarrow3p_y}$ and  $\pi \rightarrow \pi^*$ states using the aug-cc-pVDZ basis set creates 
an artificial avoided crossing away from the planar geometry that manifests as 
a pair annihilation point, where one of the states coalesces with another unphysical solution and disappears.
Since the reference SS-CASSCF\,(2,2) solution mathematically disappears, these irregularities cannot be remedied 
by post-CASSCF correlation methods such as CASPT2,\cite{Andersson1990,Andersson1992,Angeli2001} multireference CI,\cite{Werner1988} or even multi-state CASPT2.\cite{Finley1998}
Instead, a state-specific wave function approximation that is optimised in the presence of dynamic correlation
will be required to stop states from disappearing.
Therefore, there is a trade-off between coalescing solutions and root-flipping discontinuities in 
state-specific and state-averaged CASSCF, respectively

Furthermore, SS-CASSCF(2,2) calculations with the 6-31G basis set cannot capture the diffuse character of 
the $\pi \rightarrow \pi^*$  state at all, which is predicted to be too high in energy. 
This error causes an artificial avoided crossing with the $\upsigma \rightarrow\uppi^*$ excitation, 
and we were unable to find any meaningful energy surfaces for the $\uppi \rightarrow \uppi^*$ or $(\uppi)^2 \rightarrow (\uppi^*)^2$ states. 
These observations emphasise the importance of using sufficient basis sets for the excited states of interest, 
and also highlight the danger of assessing state-specific approximations using inadequate basis sets.

Ultimately, the coalescence and disappearance of solutions remains the primary obstacle to practical excited state-specific calculations.
These coalescence points are mainly due to the unbalanced description of valence and Rydberg excitations.
While this imbalance may be attributed to the lack of dynamic correlation, an alternative perspective is that the 
SS-CASSCF approximation simply is not the right reference for molecular excited states.
Since the ethylene single excitations  correspond to open-shell singlets, further restricting the wave function to  a single CSF would not change our results.
Instead, we believe that new wave function approximations, which explicitly include  
the effects of dynamic $\upsigma$-polarisation and orbital contraction in excited states, may provide more accurate
and efficient energy surfaces for photochemistry, and we intend to pursue this direction in future work.

\section*{Supporting Information}

Planar ethylene structure ($\mathrm{a_0}$) used for aug-cc-pVDZ calculations (\textcolor{blue}{xyz}).
Planar ethylene structure ($\mathrm{a_0}$) used for 6-31G calculations (\textcolor{blue}{xyz}).

\begin{acknowledgements}
H.G.A.B. gratefully acknowledges funding from New College, Oxford (Astor Junior Research Fellowship) 
and Downing College, Cambridge (Kim and Julianna Silverman Research Fellowship).
The authors thank Nick Lee, David Tew, and Pierre-Fran\c{c}ois Loos for discussions and research support.
\end{acknowledgements}

\appendix

\section{Line search at pair annihilation points}
\label{sec:linesearch}
The disappearance of a SS-CASSCF solution as the molecular structure changes indicates the existence of 
a pair annihilation point, which mathematically corresponds to the coalescence of two stationary points in a fold catastrophe.\cite{GilmoreBook} 
The Hessian index of the two solutions must differ by at most one downhill direction. 
For example, an index-1 saddle point can coalesce with a minimum or an index-2 saddle point.
At the coalescence point itself, the two solutions become identical and one of the Hessian eigenvalues becomes zero.
To find the other solution involved in this pair annihilation, we exploit the fact that the eigendirection corresponding to the zero Hessian eigenvalue  points from one solution to the other if we are close enough to the coalescence point.
We can then identify constrained stationary points of the energy using a line search along this eigendirection, and use the one that is closest to the original solution as an initial guess for a SS-CASSCF calculation. 
This subsequent SS-CASSCF state will  converge to the complementary  solution involved in the pair annihilation.
Through this procedure, we can fully map the pattern of coalescing solutions in SS-CASSCF theory.

\section*{References}
\bibliography{manuscript}

\end{document}